\documentclass[aps,prb,preprint,superscriptaddress]{revtex4-1}
\usepackage{amsmath}
\usepackage{amssymb}
\usepackage{graphicx}   
\usepackage{mhchem}
\usepackage{amsfonts}
\usepackage{tgtermes}

\begin{document}

\newcommand{\degrees}{$^\circ$C}
\newcommand{\bcao}{BaCo$_{2}$(AsO$_{4}$)$_{2}$}
\newcommand{\rucl}{$\alpha$-RuCl$_{3}$}

\title{Weak-field induced nonmagnetic state in a Co-based honeycomb}

\author{Ruidan Zhong}
\affiliation{Department of Chemistry, Princeton University, Princeton NJ 08544, USA}
\author{Tong Gao} 
\author{Nai Phuan Ong}
\affiliation{Department of Physics, Princeton University, Princeton NJ 08544, USA}
\author{Robert J. Cava}
\affiliation{Department of Chemistry, Princeton University, Princeton NJ 08544, USA}

\date{\today}

\begin{abstract}
Layered honeycomb magnets are of interest as potential realizations of the Kitaev quantum spin liquid (KQSL), a quantum state with long-range spin entanglement and an exactly solvable Hamiltonian. Conventional magnetically ordered states are present for all currently known candidate materials, however, because non-Kitaev terms in the Hamiltonians obscure the Kitaev physics. Current experimental studies of the KQSL are focused on 4$d$- or 5$d$-transition-metal-based honeycombs, in which strong spin-orbit coupling can be expected, yielding Kitaev interaction that dominate in an applied magnetic field. In contrast, for 3$d$-based layered honeycomb magnets, spin orbit coupling is weak and thus Kitaev-physics should be substantially less accessible. Here we report our studies on \bcao, for which we find that the magnetic order associated with the non-Kitaev interactions can be fully suppressed by a relatively low magnetic field, yielding a non-magnetic material and implying the presence of strong magnetic frustration and weak non-Kitaev interactions.  

\end{abstract}

\maketitle

\section*{Introduction}

Unlike the quantum spin liquids (QSL) found in geometrically frustrated quantum magnets, the Kitaev QSL arises from strong anisotropy and bond-dependent interactions that frustrate the spin configuration on a single site of a honeycomb lattice (1). The Kitaev model, which is an exactly solvable model of honeycomb lattice magnetism, has attracted considerable recent attention as it gives rise to quantum and topological spin liquids and emergent Majorana quasiparticles (1). In real materials, the spin Hamiltonian for such systems can be expressed by the sum of three terms, with $J, K$ and $\Gamma$ representing Heisenberg ($J$), Kitaev ($K$) and bond-dependent off-diagonal exchange interactions ($\Gamma$), respectively. This is known as the extended Kitaev-Heisenberg (EKH) quantum spin model (2, 3). 

In order to approach the ideal KQSL state, Kitaev interaction are required to dominate the spin Hamiltonian (4, 5). Such bond-dependent anisotropic Kitaev-type interactions are believed to dominate in materials with strong entanglement due to spin-orbit coupling (SOC) (6), and thus so far most theoretical and experimental investigations of the Kitaev quantum spin liquid state have been devoted to candidates with 4$d$- and 5$d$-transition-metal-based honeycomb lattices, including $\alpha$-RuCl$_{3}$, $A_{2}$IrO$_{3}$ ($A$=Li, Na), and H$_{3}$LiIr$_{2}$O$_{6}$ (7-10). Even though Kitaev interaction are supposed to be strong for these materials, they are nonetheless not strong enough to stabilize the QSL state. Instead, the inevitable non-Kitaev interactions present in all these systems induce conventional magnetic order at finite temperatures (11-13) obscuring the signature (e.g. a half-integer quantized thermal Hall conductivity) of the Kitaev spin liquid state that is potentially present. Theoretical and experimental efforts have shown that the non-Kitaev terms can be suppressed by applying tuning parameters, such as a magnetic field (13-15), and that the ground state in that case may in fact be the exotic Kitaev QSL phase (16, 17).

In the search for the Kitaev QSL, recent theoretical studies (18, 19) have provided new ideas for extending the candidates to high-spin d$^{7}$ electron configuration systems, especially those based on the 3$d$ transition metal ion Co$^{2+}$ ($L=1, S=3/2$) (18).  As potentially interesting systems, several Co-based materials with a honeycomb crystal structure are known, such as BaCo$_{2}$(PO$_{4}$)$_{2}$ (20), \bcao\ (21-23), Na$_{3}$Co$_{2}$SbO$_{6}$ (24, 25), and Na$_{2}$Co$_{2}$TeO$_{6}$ (24, 26). All of them exhibit conventional long-range or short-range magnetic ordering at low temperature. Here, motivated by recent theoretical models, we revisit one of these Co-based honeycomb materials, \bcao, well studied by neutron scattering in the 1970s. Though characterization of its magnetism, specific heat and thermal conductivity, we find that it is an excellent candidate for the study of Kitaev physics in a 3$d$-based material. Our work on high quality single crystals shows that the magnetic susceptibility is highly anisotropic, that the application of an appropriately orientated magnetic field of weak magnitude induces two consecutive magnetic phase transitions, and that the honeycomb magnet eventually attains a low-temperature nonmagnetic state at around 0.5 T. The behavior we observe is similar to what is observed in the well-established 4$d$-based KQSL material $\alpha$-RuCl$_{3}$. Importantly, we find that the magnetic phases present are extremely sensitive to a relatively weak in-plane field compared to the heavy-transition-metal honeycombs, which is a clear sign of weak nearest-neighbor Heisenberg interactions.

\section*{Results}

\textbf{Crystal structure and anisotropic magnetic susceptibility.} \bcao\ crystallizes in the trigonal centrosymmetric space group $R$-3 (No. 148), with the lattice parameters $a = b$ = 5.00 $\AA$, $c$ = 23.49 $\AA$. Schematic plots of the crystal structure are shown in Fig. 1\textbf{A} and \textbf{B}. The material consists of Co-based magnetic honeycomb layers, packed with an ABC periodicity along the c axis (Fig. 1\textbf{A}). As shown in Fig. 1\textbf{B}, within the plane the honeycomb structure is made of edge-sharing CoO$_{6}$ octahedra. No stacking faults or twin domains are observable in this material from our single crystal X-ray diffraction characterization, owing to the fact that the magnetic honeycomb layers are stacked in 3D through ionic bonding to the Ba and AsO$_{4}$ tetrahedra. In contrast, van der Waals interlayer bonding, which is the case for $\alpha$-RuCl$_{3}$, can lead to honeycomb plane stacking faults and coexisting structural domains (27, 28). Compared to frequently studied Kitaev physics material $\alpha$-RuCl$_{3}$ then, our material is far simpler in terms of interpreting the magnetic and thermal properties. Our dark purple single crystals of \bcao\ were obtained by the flux growth method, as shown in the inset of Fig. 1\textbf{C}.  

\begin{figure*}[t]
\centering
\includegraphics[width=16cm,height=5.5cm]{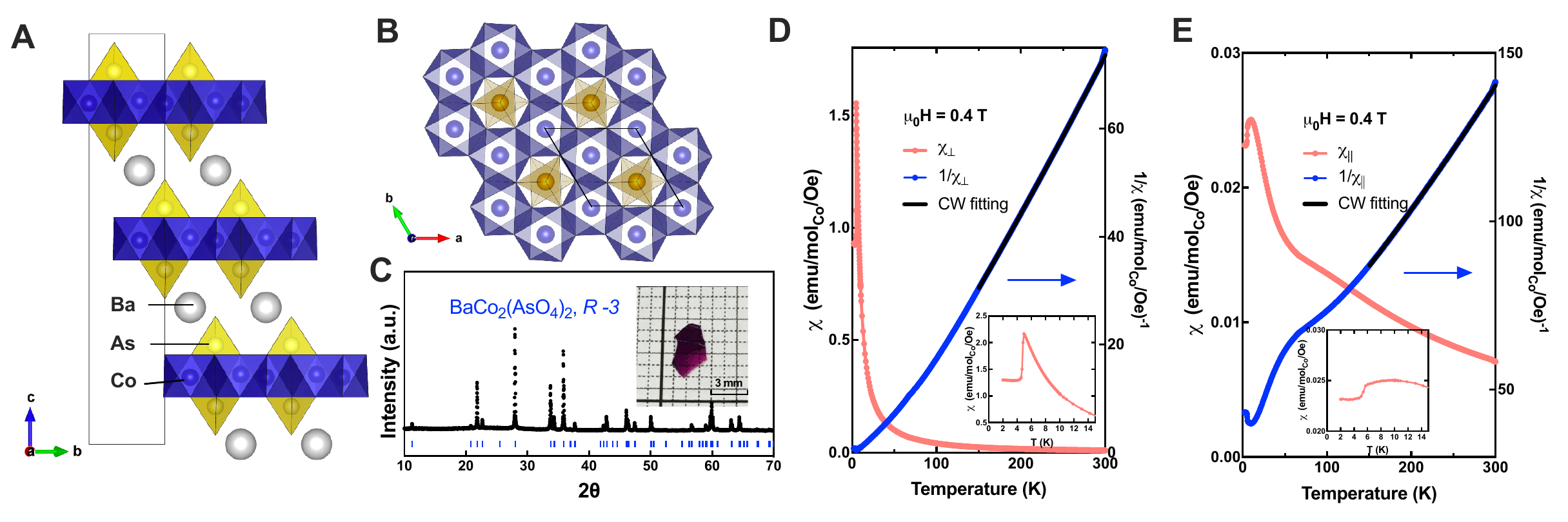}
\caption{\textbf{Crystal structure and anisotropic magnetic susceptibility.} \textbf{A} Schematic of the \bcao\ crystal structure, showing the honeycomb plane stacking along the c axis. \textbf{B}  An individual honeycomb layer made of edge-sharing CoO$_{6}$ octahedra; the [AsO$_{4}$]$^{3-}$ tetrahedra sit in the middle of each honeycomb. The lines in (\textbf{A}) and (\textbf{B}) indicate the unit cell. (\textbf{C}) Room temperature X-ray diffraction pattern of the crushed \bcao\ single crystals, indicating the high quality of the crystals. Calculated diffraction peak positions are marked by short blue ticks. The inset in (\textbf{C}) shows a photo of a dark pink single crystal. (\textbf{D, E}) dc magnetic susceptibility $\chi$ and the inverse susceptibility 1/$\chi$ as a function of temperature measured for a \bcao\ single crystal, under magnetic fields (0.4 T) applied both in-plane (\textbf{D}, H $\perp$ c) and out-of-plane (\textbf{E}, H // c). The magnetic transition is shown in detail in the insets. The Curie-Weiss fitting (black lines) results in Curie temperatures. }\label{fig:fig1}
\end{figure*}

\begin{figure*}[t]
\centering
\includegraphics[width=.7\linewidth]{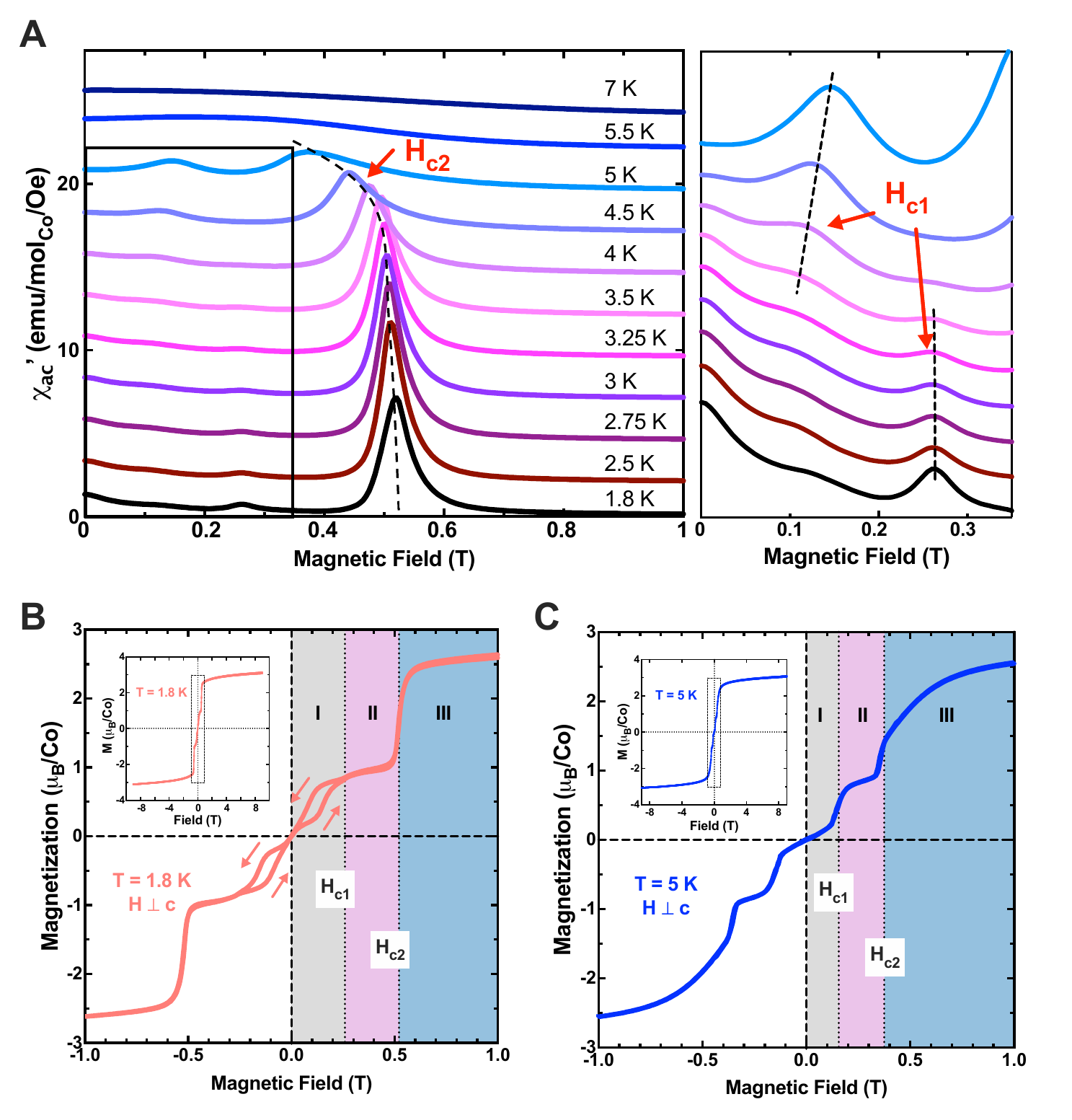}
\caption{\textbf{Field-induced phase transitions.} \textbf{A} Left panel: ac magnetic susceptibility as a function of in-plane dc field up to 1 T, measured at multiple temperatures. Dashed curve indicates the critical field H$_{c2}$ at different temperature, obtained from the peak positions of $\chi$'$_{ac}$. The box area has been zoomed in to the right panel.  Right panel: Dashed lines illustrate the critical field Hc1 at each temperature. (\textbf{B, C}) Magnetic hysteresis measured under in-plane dc field $\mid \mu_{0}H\mid$ $\leqslant$ 1 T at 1.8 K (\textbf{B}) and 5 K (\textbf{C}), respectively. The arrows in (\textbf{A}) indicate the field sweeping direction. H$_{c1}$ and H$_{c2}$ obtained in (\textbf{A}) at each temperature are indicated by the dashed lines, separating multiple magnetic phases with colorful shading. The two insets show the magnetic hysteresis at each temperature within a full field range of 9 T.}\label{fig:fig2}
\end{figure*}

\begin{figure*}[t]
\centering
\includegraphics[width=16cm,height=10.0cm]{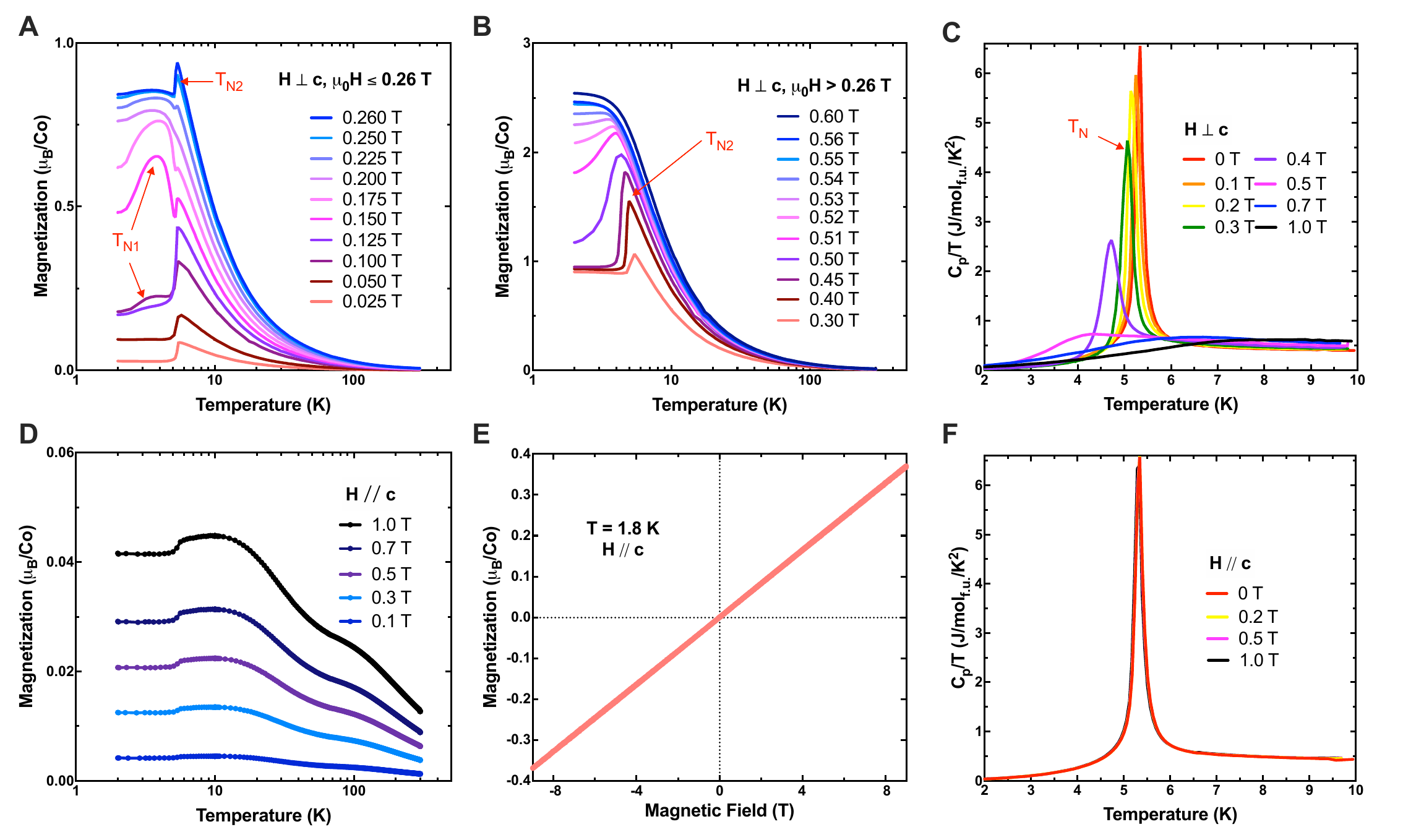}
\caption{\textbf{Weak in-plane field manipulation of the magnetic structure.} \textbf{A, B} Magnetization as a function of temperature, measured with an in-plane field (A) $\mu_{0}$H $\leqslant$ 0.26 T and (B)$\mu_{0}$H $>$ 0.26~T. $T_{N1}$ and $T_{N2}$ with red arrows indicate both phase transitions. (C) Low temperature specific heat C$_{p}$/T at several magnetic in-plane fields. $T_{N}$ with red arrow indicates the observed single peak at all temperatures associated with the first-order transition. (D) $dc$ magnetization as a function of temperature measured under several out-of-plane fields. (E) Magnetic hysteresis loop measured under out-of-plane fields up to 9 T. (F) Low temperature specific heat C$_{p}$/T measured with $dc$ fields applied perpendicular to the hexagonal plane.}\label{fig:fig3}
\end{figure*}

Anisotropic magnetization is usually found in layered magnetic oxides. To characterize the magnetic behavior of Co in the honeycomb layer of \bcao\, we measured the magnetic susceptibility of a single crystal under both in-plane and out-of-plane fields, presented in Fig. 1\textbf{D} and \textbf{E}. For magnetic fields H applied parallel to the c axis, i.e. for field perpendicular to the honeycomb plane, the spin interactions are antiferromagnetic, as evidenced by a negative Curie-Weiss (CW) temperature $\Theta_{CW, \parallel}$ = -167.7 K. However, with a field applied within the honeycomb plane, a CW temperature of $\Theta_{CW, \perp}$ = 33.8 K is derived from the fitting, indicating ferromagnetic magnetic coupling within the honeycomb planes. From these values we estimate the exchange coupling constant $J/k_{B}$ to be 3.6 K (29 and references therein). The Curie-Weiss fit of the inverse susceptibility yields effective moments of $\mu_{eff,\parallel}$ = 5.91$\mu_{B}$/Co and $\mu_{eff,\perp}$ = $\mu_{B}$/Co for H // c and H $\perp$c, respectively. Clearly these values are well above the spin-only value for the effective S = 1/2 spin configuration of Co$^{2}$+ (1.73$\mu_{B}$), due to the unquenched orbital contribution of the Co. Similarly large $\mu_{eff}$’s have been observed in other compounds consisting of Co$^{2+}$O$_{6}$ octahedra, such as Na$_{2}$BaCo(PO$_{4}$)$_{2}$  (30), which is an effective spin-1/2 system, evident by neutron scattering measurements. The strong magnetic anisotropy revealed by the susceptibility in \bcao\ is consistent with that reported for \rucl, which is believed to contribute to the anisotropic exchange interactions in the spin Hamiltonian (11, 31, 32).

\textbf{Weak field manipulation of the magnetic states.} The orientation dependent M-T data in Fig. 1\textbf{D} and \textbf{E} were obtained under a field of 0.4 T, and indicate a clear AFM transition when the field is applied in the honeycomb plane. The magnetic behavior is greatly dependent on the magnitude of the applied field, reflecting the presence of field-induced magnetic phase transitions. We have also investigated the ac susceptibility, which is a more sensitive method for determining the onset of magnetic phase transitions (33), as illustrated in Fig. 2\textbf{A}. At each constant temperature a small ac field of 5 Oe with a frequency of 5000 Hz is applied while sweeping the dc field within the honeycomb plane. At 1.8 K, two peaks in ac susceptibility are observed, corresponding to two separate magnetic phase transitions with two critical fields, $H_{c1}$ = 0.26 T and $H_{c2}$ = 0.52 T. Both critical fields are quite sensitive to temperature. The more obvious transition around 0.5 T displays a decreasing $H_{c2}$ with increasing temperature (Fig. 2\textbf{A}, left), while the one at lower field displays more complicated behavior. (In the expanded data (Fig. 2\textbf{A}, right), one can define $H_{c1}$ at two positions, within a narrow field range.) Magnetic hysteresis helps to clarify the nature of those transitions. As illustrated in Fig. 2\textbf{B}, the magnetic hysteresis loop measured at 1.8 K has a zero coercivity and a dumbbell shape. In spite of the fact that the overall magnetic interactions in this material are dominated by interplanar AFM coupling (Fig. 2\textbf{B}, inset), similar double hysteresis loop observed in the low field regime ($< H_{c1}$) have been previously reported for ferroelectrics (34), and in this case appear to result from in-plane ferromagnetic spin correlations. Previous neutron scattering studies (21, 22) reveal that the magnetic ground state of this system displays 2-dimensional spiral order in zero applied field, consisting of weakly coupled quasi-ferromagnetic in-plane chains, which may result in the dumbbell-shape hysteresis observed below $H_{c1}$. At 5 K (Fig. 2\textbf{C}), the magnetic hysteresis associated with in-plane FM correlations disappears but metamagnetic transitions at $H_{c1}$ = 0.155 T and $H_{c2}$ = 0.375 T are still observed. Thus, the field-effect on the system can be understood by two consecutive AFM-type phase transitions, at $H_{c1}$ and $H_{c2}$, resulting in three magnetic phases, as marked in Fig. 2\textbf{B} and \textbf{C}. At low temperature the in-plane ferromagnetic correlations result in magnetic hysteresis, leading to a difference in increasing and decreasing field behavior (23).

The temperature evolution of the magnetic system measured under multiple in-plane fields is illustrated in Fig. 3\textbf{A} and \textbf{B}, grouped by the field magnitude for clarity. As shown in Fig. 3\textbf{A}, at $\mu_{0}$H $\leqslant$ 0.26 T, a broad transition at lower temperature ($T_{N1}$) and a sharp AFM transition at higher temperature ($T_{N2}$) are observed (Fig. 3\textbf{A}). When $\mu_{0}$H $>$ 0.26 T, only a single AFM phase transition at $T_{N2}$ is seen. With increasing field, $T_{N2}$ gradually shifts to lower temperature and the transition is not detectable above $\mu_{0}$H $\approx$ 0.53 T (Fig. 3\textbf{B}). Similar behavior has been reported for other well-studied magnetic honeycombs such as $A_{2}$IrO$_{2}$ ($A$ = Li, Na) (13) and \rucl\ (35), in which the zigzag AFM ground state turns into a nonmagnetic state at applied fields of 4~8 Tesla. By contrast, the quantum phase transition observed in the \bcao\ happens at a much lower field (0.5 T), a reflection of relatively weaker non-Kitaev interactions in the studied compound compared to other well-studied magnetic honeycombs. 

Specific heat measurements provide an additional perspective for understanding the magnetic phase transitions. (As shown in Figure 3C, the sharp, single transition further indicates the high quality of our single crystals, which have no stacking faults.) The $\lambda$-peak associated with the long-range AFM ordering is rapidly suppressed by a relatively small in-plane-field and appears to be fully absent above 0.5 Tesla. The transition temperature $T_{N}$ observed in specific heat measurements agrees well with $T_{N2}$ obtained from magnetization, which suggests that the transition at higher temperature (~5 K) is indeed due to a magnetic phase transition that involves heat exchange. However, the broad transition $T_{N1}$ observed in magnetization does not result in a $\lambda$-shaped peak in specific heat, ruling out any typical first or second order transition. This broad transition sensitive to field may be associated with the realignment of the in-plane spins.

\begin{figure*}[t]
\centering
\includegraphics[width=7cm,height=12.0cm]{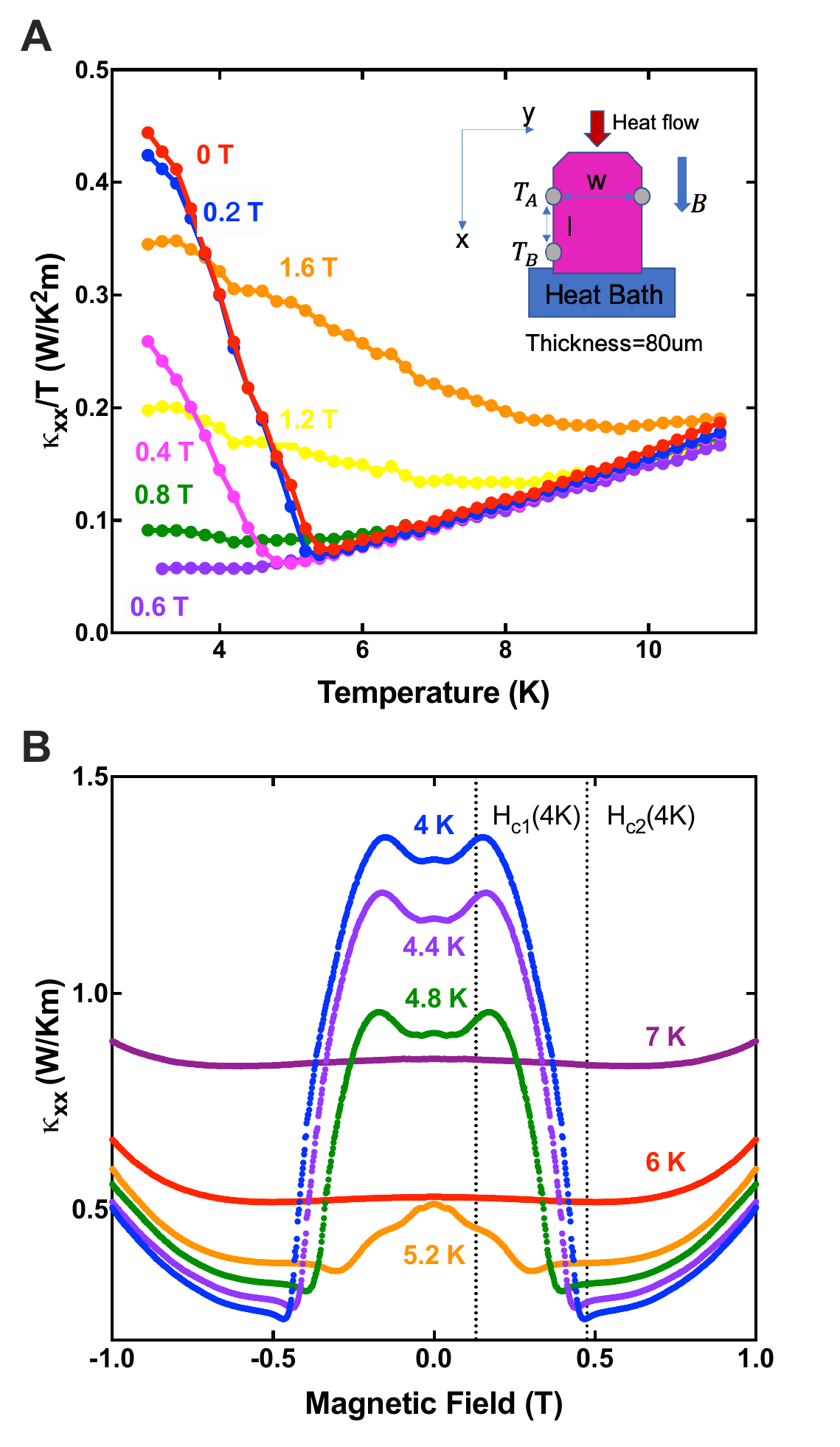}
\caption{\textbf{Temperature and field dependence of the thermal conductivity.} (\textbf{A}) Thermal conductivity over temperature$\kappa_{xx}$/T against temperature at various magnetic fields. The inset shows a schematic of the experiment setup. Two thermometers measuring temperature at A and B are marked as T$_{A}$ and T$_{B}$. External field B is applied in-plane. w and l represent the sample’s width (1.5 mm) and length (3 mm), respectively. (\textbf{B}) Thermal conductivity $\kappa_{xx}$ vs. magnetic field at various temperatures. H$_{c1}$ and H$_{c2}$ marked in the figure are obtained from ac susceptibility showing in Fig. 2\textbf{A}. The magnetic field is applied in the ab plane for all thermal conductivity measurements. }\label{fig:fig4}
\end{figure*}

In contrast to the behavior in an in-plane magnetic field, the effect of the out-of-plane field on the magnetic system is negligible. With the external field applied perpendicular to the honeycomb plane, the magnetization is identical at different fields (Fig. 3D). The corresponding magnetization shows a linear dependence on the external field (Fig. 3E), and agrees well with expectations for A-type antiferromagnetism in which the interlayer coupling is dominated by the AFM interactions. In addition, as shown in Fig. 3F, specific heat measurements reveal a single phase transition which is unaffected by an out-of-plane field up to 1 T.

\textbf{Strong temperature and field dependence of the thermal conductivity.} Thermal conductivity was also employed to characterize the honeycomb system in the vicinity of the phase transitions. Strong temperature and field dependence are observed. Representative data are shown in Fig. 4\textbf{A} and Fig. 4\textbf{B}. Fig. 4\textbf{A} shows the temperature dependence of the thermal conductivity $\kappa_{xx}$ for various in-plane applied magnetic fields. At temperatures above 5.4 K under 0 applied field, $\kappa_{xx}$/T decreases as temperature decreases. As the temperature drops below $T_{N}$ = 5.4 K, however, the material undergoes a phase transition, and a long-range AFM order starts to develop. This greatly enhances the thermal conductivity due to the suppression of the phonon-magnon scattering in the AFM ordered state, which is a common observation for magnets. This enhancement is suppressed with a magnetic field of 0.4 Tesla and can no longer be seen with an applied field of 0.6 Tesla as the field suppresses AFM order and drives the sample into nonmagnetic states. The detailed field dependence of the thermal conductivity is shown in Fig. 4\textbf{B}. There is no field dependence of thermal conductivity at temperatures above 5.4 K. At 4 K, the thermal conductivity increases slightly as the external field reaches $H_{c1}$ and decreases dramatically between $H_{c1}$ and $H_{c2}$ as the long-range order is continuously suppressed by an external in-plane field. The material eventually undergoes a metamagnetic phase transition into a nonmagnetic state as the external field goes beyond $H_{c2}$. In the field-induced nonmagnetic state, $\kappa_{xx}$ increases with the magnetic field due to increasing magnon stiffness. Spin polarization enhances as field increases, resulting in weaker spin-phonon scattering and thus larger thermal conductivity. Similar observations have reported for \rucl, in which the thermal conductivity reaches a minimum at the critical field (36) and $\kappa_{xx}$ is enhanced greatly with increasing field in the nonmagnetic state (37). The enhancement of $\kappa_{xx}$ at low temperature may be attributed to the low-energy excitations of the field-induced spin-liquid phase (38). 

\begin{figure*}[t]
\centering
\includegraphics[width=.6\linewidth]{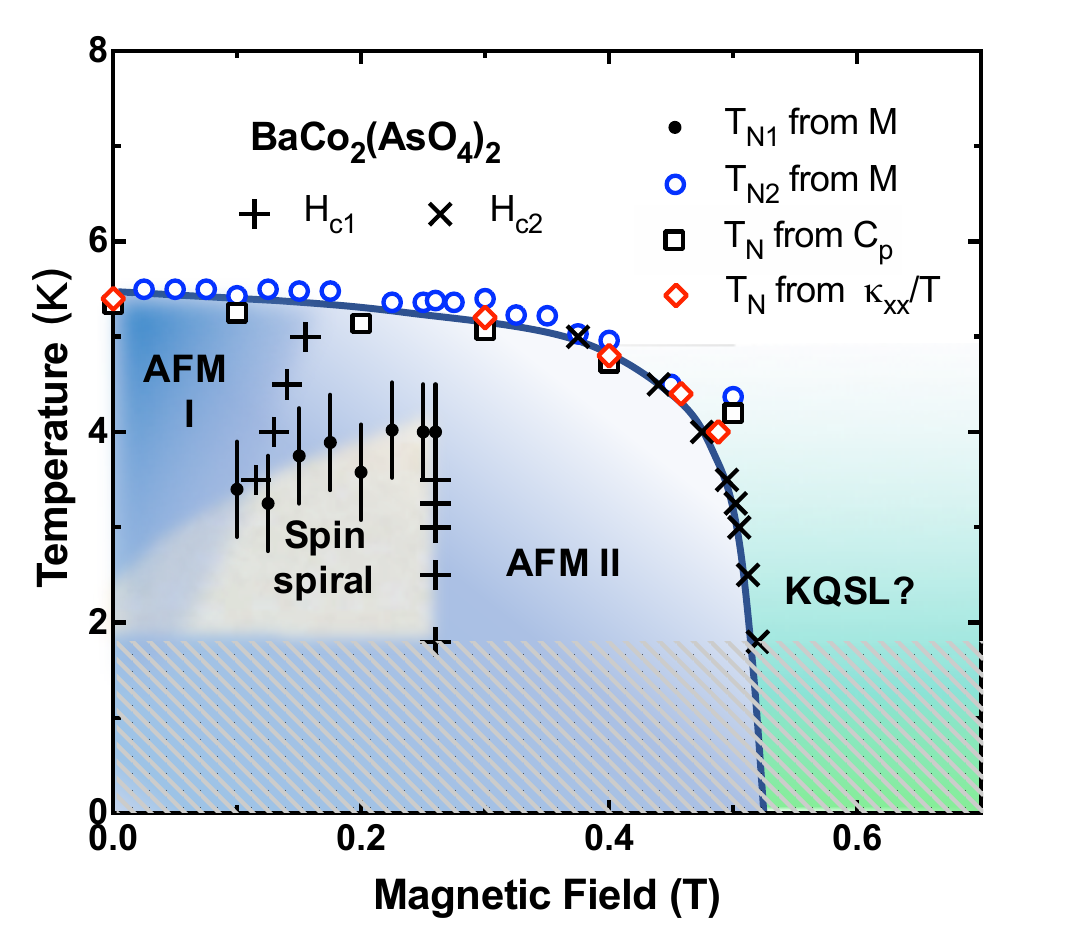}
\caption{\textbf{A phase diagram showing the evolution of the AFM order under in-plane field in \bcao.} (\textbf{A}) T$_{N1}$ (black dots) is defined as the temperature where the low-temperature broad transition has a maximum in Fig. 3\textbf{A}. The corresponding error bars illustrate the full-width-half-maximum (FWHM) of the hump. The Néel temperature T$_{N2}$ (blue circles) is determined from magnetization in Fig. 3\textbf{B}. T$_{N}$ is determined from specific heat data (black squares) showing in Fig. 3\textbf{C} or thermal conductivity measurements (red diamonds) showing in Fig. 4\textbf{A}. The boundary between different phases, H$_{c1}$ and H$_{c2}$ is determined from ac susceptibility in Fig. 2\textbf{A}. }\label{fig:fig5}
\end{figure*}

\textbf{Magnetic phase diagram.} Employing our data and data from the previous studies on the magnetic structure of \bcao, we generate a magnetic phase diagram for this honeycomb material, shown in Fig. 5. The spin spiral structure in the ground state of \bcao\ can be easily eliminated through spin realignment driven by either temperature or an in-plane magnetic field. The resulting collinear AFM state may have one of several possible magnetic structures, i.e. zigzag, Néel or stripy on the honeycomb lattice. The transition between these states at $H_{c1}$ results in the metamagnetism observed in M vs. H (Fig. 2 \textbf{C} and \textbf{D}). Such a process involves no change in order parameter thus no peak was observed in the specific heat measurements. The detailed magnetic structures of the AFM phases I and II should be further clarified by neutron scattering. Eventually the in-plane field suppresses the conventional Heisenberg interactions and drives the system to become a spin-liquid state at $T_{N2}$ ($T_{N}$) and $H_{c2}$.

Similar observations have been reported for \rucl\ (39-41), which show that the zigzag order within the honeycomb planes is continuously suppressed with increasing in-plane field until completely being washed out around 8 T. Supported by the nuclear magnetic resonance (NMR) (13,37), thermal Hall conductivity (42) and inelastic neutron scattering (16, 17) measurements on that material, the field-induced nonmagnetic phase is attributed to the Kitaev quantum spin liquid state. Other Kitaev systems are also revealed to be excellent candidates for field-induced quantum spin liquids, such as Na$_{2}$IrO$_{3}$ (43) and $\beta$-Li$_{2}$IrO$_{3}$ (13). Compared to the known magnetic honeycomb materials, \bcao\ exhibits similar effects, although at a much weaker magnitude of applied field that is easily accessible in most experimental approaches. The phenomenological similarities in \bcao\ implies that the field-induced spin-liquid phase may dominated by the Kitaev interaction as well, leading to an excitement on further determinations of the possible Kitaev quantum spin liquid state.

The connection between the magnetic field sensitivity and non-Kitaev (Heisenberg and other) interactions is evident. In the Ru- or Ir-honeycombs, even with strong SOC and prominent Kitaev term in the spin interactions, the Kitaev physics is not sufficient to stabilize the KQSL ground state, due to the existence of the strong Heisenberg interactions. Thus, extra tuning parameters such as external fields that kill these non-trivial interactions are necessary in order to realize the pure Kitaev physics. In \bcao, in contrast, we find a complex multi-phase transition and that the magnetic ordering can be totally washed out by a weak field of $\sim$ 0.5 T, indicating the presence of weak nearest-neighbor Heisenberg and other non-Kitaev interactions, likely due to strong magnetic frustration.

\section*{Discussion}
Since the two essential elements for the Kitaev interaction are the bond-dependent exchange coupling and its anisotropy, honeycomb materials based on the 3$d$ transition metal Co$^{2+}$ are good candidates; the orbital degeneracy derived from the unquenched orbital contribution to the magnetic moment is responsible for the interaction anisotropy (18, 19). In addition, the relativistic spin-orbit coupling $\lambda$ is large compared to the Jahn-Teller (JT) coupling EJT and exchange interaction J, giving rise to the relatively large SOC in 3$d$ transition metal Co compounds (14). Compared to the widely studied low-spin d$^{5}$ ions, the additional ferromagnetic spin exchange of the eg electrons in the d$^{7}$ configuration (i.e. as seen for Co$^{2+}$) may largely compensate the antiferromagnetic contribution of the Heisenberg term (18) in the Hamiltonian, leading to a proximate Kitaev QSL state. Therefore, in addition to the widely discussed heavy transition metal systems based on Ir and Ru, Co-based honeycombs can also be promising candidates in the search for the KQSL state. 

Thus, cobalt honeycomb lattice materials may host Kitaev interaction. However, they are not necessarily dominated by Kitaev interaction under a magnetic field as the conventional Heisenberg term in the Hamiltonian is usually dominant. Here, however, we characterize a Co-based honeycomb system that displays interesting behavior under an appropriately oriented magnetic field. With a rather weak field applied in the honeycomb plane, the magnetic ordering melts to a disordered state, evident by magnetization, specific heat and thermal conductivity measurements, summarized in the H-T phase diagram, Fig. 5. In Co-based transition metal compounds with residual orbital angular momenta, the large degeneracy of the unquenched orbital leads to frustrating interactions and anisotropic magnetic properties, which results in multiple competing phases with similar energy scale (44). This is further proved by the observed sensitivity of such competing phases to the field. 

In this work, we study a known Co-based honeycomb material, \bcao, from the perspective of Kitaev physics by performing measurements on high quality single crystals. We observe a field-tuned low-temperature spin-liquid state, which shows similar behavior to \rucl\ and other Kitaev systems with regard to magnetization, specific heat and thermal conductivity, making this system an excellent candidate for a field-induced quantum spin liquid state. These features observed in a 3$d$-element-based honeycomb extend the current interest on extensively studied heavy-element systems into a new regime. 

Our results are strong support for the theoretical predictions about the possibility to realize a KQSL in a Co-based honeycomb material. Importantly, the relatively weak critical field of $\sim$ 0.5 T and the availability of single crystals will motivate wider studies. To clarify the nature of the field-induced nonmagnetic state, we hope that this study will inspire the search for the signature of the KQSL state - like half-integer quantized thermal Hall conductivity and fractionalized spin excitations in inelastic neutron scattering.

\section*{Materials and methods}

\subsection*{Crystal growth}
Due to the toxicity and low boiling point of the starting material As$_{2}$O$_{5}$, a powder sample of \bcao\ was prepared by solid state reaction before the single crystal growth by the flux method. Stoichiometric mixtures of powder of BaO (99.9\%), CoO (99\%), As2O5 (99.99\%) were ground and packed in to an alumina crucible, and the crucible was then carefully sealed in a quartz tube. The whole process was performed in an inert gas glove box to avoid the inhalation of a toxic substance as well as the deterioration of air-sensitive BaO. The sealed starting materials were heated at 300 \degrees for 12 hrs and then heated to 850 \degrees for 24 hrs. The obtained powder was then mixed and ground well with flux medium NaCl in a molar ratio of 1:5. The mixed materials were loaded into a capped alumina crucible, kept at 900 \degrees for 2 hrs and then slowly cooled down to 700 \degrees in a rate of 3\degrees /hr. Then the whole furnace was quickly cooled down to room temperature. The crystals can be separated from the flux by dissolving in hot water. 

\subsection*{Magnetization and thermodynamic measurements }
The dc, ac magnetic susceptibility and specific heat were measured on single crystals in a physical property measurement system (PPMS) that cooled to 1.8 K (PPMS-DynaCool, Quantum Design), equipped with a vibrating sample magnetometer (VSM) option. Heat capacity measurements were conducted on a Quantum Design PPMS Evercool II with applied magnetic field up to 1 T. All measurements were carried out with a designed crystal orientation. 

\subsection*{Thermal conductivity}
For thermal transport measurements, we reduced the sample thickness to 80 microns by polishing with diamond paper. The sample was of approximate size 3*4*0.08 mm. Lakeshore Ruthenium oxide thermometers (Rx-102) were employed for negligible field-induced corrections. The thermometers were attached to the sample via two thick gold wires with silver paint in order to provide a good thermal connection. The thermometer resistance was measured in a four-wire geometry with phosphor bronze wire in order to minimize the heat leak. The sample chamber was evacuated below 10$^{-6}$ mbar. The field sweep rate was limited to 0.02 T/min to minimize the eddy current heating and magnetocaloric effect. The experiments were conducted in two ways: temperature scanning with fixed field and field sweeping with constant heating power.

\begin{acknowledgments}
\textbf{Funding}: The materials synthesis, magnetic and thermodynamic characterization was supported by the Gordon and Betty Moore EPiQS program, grant GBMF-4412. The thermal conductivity research was supported by the Department of Energy (grant DE-SC0017863) and a US National Science Foundation MRSEC award (grant DMR 1420541). 

\textbf{Author contributions}: R.C. and N.P.O. conceived of the experiment. R.Z synthesized the materials. R.Z. and T.G. performed the experiments and analyzed the data. All authors contributed to the writing of this paper. 

\textbf{Competing interests}: The authors declare no competing interests.

\textbf{Data and materials availability}: All data supporting the stated conclusions of the manuscript are in the paper. Additional data are available from the authors upon request.

\end{acknowledgments}

\end{document}